\documentclass[amsmath,amssymb,superscriptaddress,aps,prl,reprint, 10pt]{revtex4-2}

\usepackage{layouts}
\usepackage[english]{babel}
\usepackage[utf8]{inputenc}
\usepackage[T1]{fontenc}                        
\usepackage{graphicx}   
\usepackage{sidecap} 
\sidecaptionvpos{figure}{t}                        
\usepackage[section,below]{placeins}            
\usepackage{mwe}
\usepackage{bm}
\usepackage{mathtools}                          
\usepackage{physics}
\allowdisplaybreaks                             
\usepackage{hyperref}
\hypersetup{hidelinks,
            colorlinks=true,
            allcolors=[RGB]{0,84,159}}

\usepackage[capitalise]{cleveref}               
\crefname{fig_a}{Fig.}{Fig.}                    
\crefname{fig_b}{Fig.}{Fig.}                    
\crefname{fig_c}{Fig.}{Fig.}                    
\Crefname{fig_a}{Figure}{Figure}                
\Crefname{fig_b}{Figure}{Figure} 
\Crefname{fig_c}{Figure}{Figure} 
\creflabelformat{fig_a}{#2{#1(a)}#3}
\creflabelformat{fig_b}{#2{#1(b)}#3}
\creflabelformat{fig_c}{#2{#1(c)}#3}

\newcommand*\subtxt[1]{_{\textnormal{#1}}}
\DeclareRobustCommand\_{\ifmmode\expandafter\subtxt\else\textunderscore\fi}
\newcommand*\supertxt[1]{^{\textnormal{#1}}}
\DeclareRobustCommand\^{\ifmmode\expandafter\supertxt\else\textasciicircum\fi}

\newcommand{\vast}{\bBigg@{4}}
\newcommand{\Vast}{\bBigg@{5}}

\newcommand{\affiliationRWTH}{
Institut f\"ur Theorie der Statistischen Physik, RWTH Aachen University and JARA-Fundamentals of Future Information Technology, 52056 Aachen, Germany
}
\newcommand{\affiliationMPSD}{
Max Planck Institute for the Structure and Dynamics of Matter,
Center for Free-Electron Laser Science (CFEL),
Luruper Chaussee 149, 22761 Hamburg, Germany
}

\begin{document}

\title{Dynamical Onset of Light-Induced Unconventional Superconductivity -- \\ a Yukawa-Sachdev-Ye-Kitaev study}

\author{Lukas Grunwald}
\email{lukas.grunwald@mpsd.mpg.de}
\affiliation{\affiliationRWTH}
\affiliation{\affiliationMPSD}

\author{Giacomo Passetti}
\affiliation{\affiliationRWTH}

\author{Dante M.~Kennes}
\email{dante.kennes@mpsd.mpg.de}
\affiliation{\affiliationRWTH}
\affiliation{\affiliationMPSD}

\date{\today}

\begin{abstract}
We investigate the dynamical onset of superconductivity in the exactly solvable Yukawa-Sachdev-Ye-Kitaev model. 
It hosts an unconventional superconducting phase that emerges out of a non-Fermi liquid normal state, providing a toy model for superconductivity in a strongly correlated system.
Analyzing dynamical protocols motivated by theoretical mechanisms proposed for light-induced superconductivity, that is light-induced cooling and the dressing of Hamiltonian parameters, we investigate the exact relaxation resulting out of undercooling and interaction quenches.
While, in contrast to BCS theory, it is not possible for superconductivity to emerge following interaction quenches across the superconducting phase transition, we find that the dynamical relaxation of undercooled states universally leads to superconductivity. Despite the strong correlations, the emerging order parameter dynamics are well captured by a coarse grained Ginzburg-Landau theory for which we determine all parameters from microscopics.
\end{abstract} 

\maketitle

The irradiation of matter with light has emerged as a powerful tool for controlling quantum materials \cite{delatorreColloquiumNonthermal2021,kennesNewEra2022,basovPropertiesDemand2017,disaEngineeringCrystal2021,tokuraEmergentFunctions2017}. 
Amongst the plethora of tantalizing light-induced phenomena, one of the most striking instances
is light-induced superconductivity, in which superconductor-like behavior is induced after photo-excitation at temperatures larger than the equilibrium transition temperature \cite{cavalleriPhotoinducedSuperconductivity2018}. 
This effect has been experimentally observed in many systems including cuprates \cite{huOpticallyEnhanced2014,mankowskyNonlinearLattice2014,faustiLightInducedSuperconductivity2011,kaiserOpticallyInduced2014,nicolettiOpticallyInduced2014,creminPhotoenhancedMetastable2019}, iron-based superconductors \cite{suzukiPhotoinducedPossible2019}, fullerides \cite{mitranoPossibleLightinduced2016,cantaluppiPressureTuning2018,roweGiantResonant2023} and organic superconductors \cite{buzziPhotomolecularHighTemperature2020}, but the underlying mechanism leading to the dynamical formation of Cooper pairs is still highly debated. 

The theoretical proposals can be broadly classified into two groups. One class of interpretations relies on a light-induced modification of Hamiltonian parameters (dressed Hamiltonians), leading to static and dynamic changes of the free energy landscape 
\cite{kennesTransientSuperconductivity2017,knapDynamicalCooper2016,babadiTheoryParametrically2017,murakamiNonequilibriumSteady2017,dasariTransientFloquet2018,sentefLightenhancedElectronphonon2017,komnikBCSTheory2016,sentefTheoryLightenhanced2016,murakamiInteractionQuench2015,eckhardtTheoryResonantly2023,kimEnhancingSuperconductivity2016,mazzaNonequilibriumSuperconductivity2017,daiSuperconductinglikeResponse2021,kennesLightinducedWave2019,patelLightinducedEnhancement2016,rainesEnhancementSuperconductivity2015,tindallDynamicalOrder2020,idoCorrelationinducedSuperconductivity2017,dolgirevPeriodicDynamics2022,chattopadhyayMechanismsLongLived2023}, 
while the second group attributes the emergent superconducting behavior, to effective light-induced cooling (dressed statistics) 
\cite{dennyProposedParametric2015,navaCoolingQuasiparticles2018,wernerEntropycooledNonequilibrium2019,wernerLightinducedEvaporative2019,robertsonNonequilibriumEnhancement2009,robertsonDynamicStimulation2011,coulthardEnhancementSuperexchange2017,okamotoTheoryEnhanced2016}. Both schemes might be distinguishable by the dynamical formation of superconductivity following the irradiation with light, calling for detailed analysis of the real time dynamics.

Strong electronic correlations, especially in cuprates and iron based superconductors, make systematic theoretical investigations challenging. In these systems, superconductivity is believed to emerge out of a strongly correlated non-Fermi liquid (nFl) normal state \cite{sachdevQuantumPhase2011,kasaharaEvolutionNonFermi2010,hashimotoSharpPeak2012,sigristPhenomenologicalTheory1991} and studies of the real time emergence of the superconducting condensate typically have to rely on perturbative approximations or purely phenomenological models. It is hence highly desirable to investigate the dynamical onset of light-induced superconductivity in a strongly correlated, yet theoretically controllable system, to gain intuition and assess the validity of common approximations.

%
\begin{figure}[t]
    \centering
    \includegraphics[width = \columnwidth]{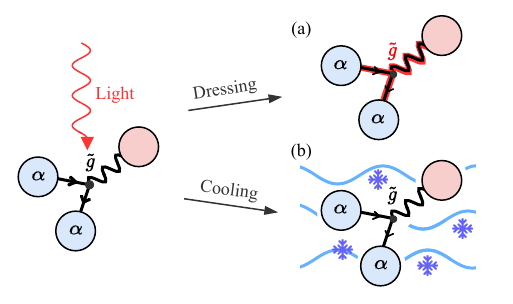} 
    \caption{\textbf{Cartoon of light irradiated YSYK model.} Irradiation of the YSYK model consisting of randomly interacting fermions (blue) and bosons (red) could dynamically induce superconductivity due to two effective mechanisms. (a) Dressing of coupling constants; here modeled as interaction quenches. (b) Cooling of the system; here modeled as the relaxation of an undercooled normal state into the superconducting ground state.}
    \label{fig:cartoon}
\end{figure}
In this letter we investigate the dynamical onset of superconductivity in the Yukawa-Sachdev-Ye-Kitaev (YSYK) model
\cite{esterlisCooperPairing2019, wangSolvableStrongCoupling2020, choiPairingInstabilities2022, classenSuperconductivityIncoherent2021, davisQuantumChaos2022, hauckEliashbergEquations2020, inkofQuantumCritical2021, panYukawaSYKModel2021, valentinisBCSIncoherent2023, valentinisCorrelationPhase2023, wangPhaseDiagram2021, wangQuantumPhase2020},
which is a solvable toy model for strongly correlated superconductivity. It presents a fermion-boson generalization of the fermionic Sachdev-Ye-Kitaev (SYK) model \cite{ sachdevGaplessSpinfluid1993,kitaevSimpleModel2015,chowdhurySachdevYeKitaevModels2022} and provides the unprecedented possibility to study the interplay of strong correlations and superconductivity in non-equilibrium without any approximations. The YSYK model consists of dispersionless bosons, randomly interacting with spinful fermions and hosts an unconventional superconducting state that emerges out of a non-Fermi liquid normal state, reminiscent of the situation observed in cuprates and iron-based superconductors \cite{sachdevQuantumPhase2011,kasaharaEvolutionNonFermi2010,hashimotoSharpPeak2012,sigristPhenomenologicalTheory1991}.

Extending previous non-equilibrium SYK studies 
\cite{alexandrovOutofequilibriumPhenomena2023, cheipeshQuantumTunneling2021, eberleinQuantumQuench2017, esinUniversalTransport2022, guoTransportChaos2019, haldarNumericalStudy2023, haldarQuenchThermalization2019, jhaChargeDynamics2023, kuhlenkampPeriodicallyDriven2020, kulkarniSYKLindbladian2021, larzulAreFast2022, larzulEnergyTransport2022, larzulQuenchesPre2022, ohanesjanEnergyDynamics2022, rossiniQuantumAdvantage2020, zanociNearEquilibriumApproach2022, zhangEvaporationDynamics2019,louwThermalizationMany2022},
to the YSYK model by include bosons and superconductivity, we investigate two dynamical protocols motivated by the theoretical proposals for mechanisms of light-induced superconductivity [\cref{fig:cartoon}]. 

First, we consider (quasi-static) light-induced modifications of Hamiltonian parameters, that we model as interaction quenches across the superconducting phase transition. We find, that in contrast to BCS theory \cite{barankovCollectiveRabi2004} and previous studies in the Hubbard model
\cite{bauerDynamicalInstabilities2015,
ecksteinInteractionQuench2010,
stahlElectronicFluctuation2021,
tsujiNonthermalAntiferromagnetic2013,
wernerNonthermalSymmetrybroken2012},
it is not possible to induce superconductivity via interaction quenches in the YSYK model due to excessive heating of the nFl normal state.
%
The light-induced cooling on the other hand, which we model as the relaxation of undercooled normal states to the superconducting ground state, is found to universally lead to superconductivity at late times.
Providing a full characterization of the non-equilibrium relaxation, we identify an overdamped and oscillatory relaxation regime, and demonstrate that a coarse grained Ginzburg Landau theory can be effectively determined from microscopics to describe the emerging order parameter dynamic despite the strong correlations. 

\paragraph{\textbf{Model} ---}
The YSYK model describes $N$ fermions randomly interacting with $N$ dispersionless bosons and is defined by the Hamiltonian \cite{esterlisCooperPairing2019,wangSolvableStrongCoupling2020,choiPairingInstabilities2022,hauckEliashbergEquations2020}
\begin{equation}
    H = \frac{1}{2}\sum_{k=1}^N \big( \pi_k^2 + \omega_0^2 \phi_k^2 \big)
    + \frac{1}{N}\sum_{ijk}^N\sum_{\alpha = \pm} g_{ij,k} \phi_k \, 
    c_{i\alpha}^\dagger c_{j\alpha}.
    \label{eq:ysyk_hamiltonian}
\end{equation}
Here $c^{(\dagger)}_{i\alpha}$ are fermionic ladder operators, while the canonical boson fields $\phi_k$ and their conjugate momentum $\pi_k$ satisfy $[\phi_k, \pi_{k'}] = i\delta_{k k'}$. The electron-boson vertices $g_{ij,k}$ are uncorrelated, random all-to-all couplings which are sampled from a Gaussian-orthogonal ensemble (GOE) \cite{akemannOxfordHandbook2015} with mean $\overline{g_{ij,k}} = 0$ and variance $\overline{ g_{ij,k}g^*_{ij,k}} = g^2$. 

The normal state at low temperatures, is a strongly interacting quantum critical non-Fermi liquid (nFl) with power law fermionic $G(\omega) \sim \abs{\omega}^{2\Delta - 1}$ and bosonic $D(\omega) \sim \abs{\omega}^{1 - 4\Delta}$ spectral functions and universal $\Delta \approx 0.42$ \cite{esterlisCooperPairing2019,classenSuperconductivityIncoherent2021}. Allowing for pairing (U(1) symmetry breaking), superconductivity emerges for all coupling strengths $g$. It supersedes the quantum critical nFl and is the final low temperature state. The superconducting condensate is a strongly interacting Cooper-pair fluid, unlike the non-interacting pair fluid in BCS theory \cite{tinkhamIntroductionSuperconductivity2004}. The YSYK model hence provides a paradigmatic example for strong coupling unconventional superconductivity.

\paragraph*{\textbf{Non-equilibrium formalism} ---}
In the thermodynamic limit $N \to \infty$ the model becomes self-averaging
\footnote{Self-averaging in this context means that each disorder realization is equivalent to the averaged theory, as well as that all Greens functions are diagonal in the site indices and the same on all sites.}
and exactly solvable due to the vanishing of vertex corrections. This allows the derivation of a closed system of self-consistent equations for fermion, boson and anomalous Green's functions (GF) on the Keldysh contour. Focusing on s-wave superconductivity, we introduce Nambu vectors $\psi_i = (c_{i\uparrow}, c^\dagger_{i \downarrow})\^T$ and the disorder-averaged Nambu-Keldysh $\boldsymbol{G}(t,t')$ and bosonic $D(t,t')$ GF's defined by
\begin{align}
    \boldsymbol{G}(t,t') &= -i/N \sum_i \overline{\expval*{T_\mathcal{C} \big[ \psi_i(t) \psi^\dagger_i(t') \big] }}, \\
    D(t, t') &= -i/N \sum_k \overline{\expval{T_\mathcal{C} \big[\phi_k(t) \phi_k(t') \big]}},
\end{align}
where $T_\mathcal{C}$ denotes contour-time ordering along the Keldysh contour $\mathcal{C}$ \cite{rammerQuantumField2007}. The time evolution of the Greens functions is governed by the Kadanoff-Baym equations (KBe)
\begin{align}
    \label{eq:kbe_fermionic}
    \!\! \boldsymbol{G}(t,t') &= \boldsymbol{G}_0(t, t')
     + \! \int_\mathcal{C} \dd{1} \! \dd{2} \boldsymbol{G}_0(t, 1) \boldsymbol{\Sigma}(1, 2) \boldsymbol{G}(2, t'), \\
    \!\! D(t,t') &= D_0(t, t') 
    + \int_\mathcal{C} \dd{1} \! \dd{2} D_0(t, 1) \Pi(1, 2) D(2, t'),
    \label{eq:kbe_bosonic}
\end{align}
with exact fermionic and bosonic self-energies \cite{supplementary}
\begin{align}
    \label{eq:ysyk_fermionic}
    \boldsymbol{\Sigma}(t,t') &= i g^2 D(t,t') 
    \left[\sigma_3 \boldsymbol{G}(t,t') \sigma_3 \right], \\
    \Pi(t,t') &= -ig^2 \tr 
    \left[\sigma_3 \boldsymbol{G}(t,t') \sigma_3 \boldsymbol{G}(t',t) \right].
    \label{eq:ysyk_bosonic}
\end{align}
\Crefrange{eq:kbe_fermionic}{eq:ysyk_bosonic} represent the exact solution of the YSYK model, and they are structurally similar to the famous Migdal-Eliashberg equations of superconductivity \cite{ummarinoEliashbergTheory2013}, but on the Keldysh contour. Providing initial conditions in the lower quadrant of the two-time plane ($t,t' < 0$), we start the non-equilibrium protocol at $t = 0$ and follow the exact evolution of the system by integrating the resulting equations numerically; see Supplementary Material for details \cite{supplementary}.
%

After the full integration of the KBe's, the dynamics of the superconducting order parameter
\begin{equation}
    \Delta(t) = \frac{1}{N} \sum_i \overline{\abs{\expval{c_{i\uparrow}(t) c_{i\downarrow}(t)}}},
    \label{eq:ysyk_gap_function}
\end{equation}
can be directly extracted from the GF's and provides a measure for the pairing strength of the system.
Using generalized fluctuation dissipation relations (gFDR), we can further extract effective temperatures and non-equilibrium occupation functions. To that end, we transform the GF's to center of mass coordinate $\mathcal{T} = (t + t') / 2$ and relative coordinate $\tau = t - t'$ and Fourier transform the latter (Wigner transformation). Writing the gFDR 
\begin{equation} 
    G\^K(\mathcal{T}, \omega) = 2i \big(1 + 2 n(\mathcal{T}, \omega) \big)
    \Im G\^R(\mathcal{T}, \omega),
     \label{eq:ysyk_Teff}
\end{equation}
with $G\^{R/K}$ the normal state electronic GF's, defines the electronic non-equilibrium occupation function $n(\mathcal{T}, \omega)$, which in thermal equilibrium is a Fermi-distribution $n\_f = (e^{\beta \omega} + 1)^{-1}$. In non-equilibrium, we can extract an effective temperature $T\_{eff}(\mathcal{T})$ from \cref{eq:ysyk_Teff}, provided that $n(\mathcal{T}, \omega)$ resembles $n\_f$ at low energies \cite{kuhlenkampPeriodicallyDriven2020}.

To study the dynamical transition from a normal into a superconducting state, we provide an explicit U(1) symmetry breaking at $t = 0$, which is implemented as
\begin{align}
    \mathcal{H}(t) = H\_{YSYK} + \alpha \theta(t) \sum_i 
    \left(
        c_{i\uparrow} c_{i \downarrow} + \textnormal{h.c.}
    \right).
    \label{eq:symmetry_breaking}
\end{align}
The symmetry breaking strength $\alpha$, provides an infinitesimal seed to start the relaxation process and is chosen as to not affect the final state after relaxation \cite{supplementary}.

\paragraph*{\textbf{Dressed Hamiltonians} ---}

\begin{figure}[t]
    \centering
    \includegraphics[width = \columnwidth]{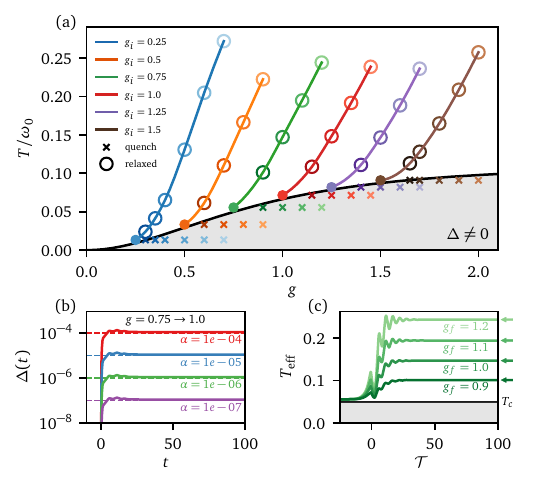}
    \caption{\textbf{Quenches across superconducting transition.} 
    (a) 
    Starting from a normal state $T\_i > T\_c(g\_i)$ (full circles) we quench into the superconducting phase such that $T\_i < T\_c(g\_f)$ (crosses). In the long time limit, the system always heats to $T\_{eff}(\infty) > T\_c(g\_f)$ (open circles), prohibiting the emergence of superconductivity.
    (b) Order parameter $\Delta(t)$ [\cref{eq:ysyk_gap_function}] for various symmetry breaking strengths $\alpha$. We observe $\Delta(t) \sim \mathcal{O}(\alpha)$, showing that no transient superconductivity emerges. (c) Effective Temperature $T\_{eff}$ extracted from gFDR for $g\_i = 0.75$, together with $T\_{inst}$ (color matched arrow), illustrating thermalization at late times.}
    \label{fig:quench_phasediagram}
    \label[fig_a]{fig:quench_phase}
    \label[fig_b]{fig:quench_gap}
    \label[fig_c]{fig:quench_temp}
\end{figure}

First, we consider light-induced modifications of Hamiltonian parameters, that we model as interaction quenches across the superconducting phase transition.
The quenches are implemented as a time dependent fermion-boson coupling
\begin{align*}
    H\_{int} \sim f(t) g_{ij,k} \phi_k \, 
    c_{i\alpha}^\dagger c_{j\alpha}, \quad \text{with } f(t) = g\_i \theta(-t) + g\_f \theta(t),
\end{align*}
that modifies the self-energies \cref{eq:ysyk_fermionic,eq:ysyk_bosonic} via $g^2 \to g^2 f(t)f(t')$. Starting from a normal state at $g_i$ with temperature $T\_i > T\_c(g\_i)$ we quench into the superconducting phase $g\_i \to g_f$ such that $T\_i < T\_c(g\_f)$. The exact dynamics following the interaction quench is illustrated in \cref{fig:quench_phasediagram}.
%

Shortly after the quench, the system is non-thermal and observables show an oscillatory dynamics [\cref{fig:quench_temp}], with amplitudes that increase in the strong coupling regime, similar to observations made for quenches in the Holstein model \cite{murakamiInteractionQuench2015}. At larger times, the system universally relaxes towards a thermalized equilibrium state. The final temperature is determined from matching the instant energy density $\epsilon(t) = \expval{H\_{YSYK}(t)}$ with that of a thermal ensemble $\expval{\dots}_{T\_{inst}}$ by solving the self-consistency relation
%
$
    \epsilon(t) = \expval{H(t)}_{T\_{inst}(t)},
$
%
hence defining the instant temperature $T\_{inst}(t)$. For the current protocol $T\_{inst}(t)$ is constant for $t \gtrless 0$ respectively and agrees with $T\_{eff}$ in the long time limit [\cref{fig:quench_temp}], indicating thermalization \cite{kuhlenkampPeriodicallyDriven2020}. This we confirm by also analyzing other thermodynamic observables and non-equilibrium occupation functions. The final temperature after the quench $T\_f = T\_{eff}(t \to \infty)$ is determined by a power law $(T\_i - T\_f) \sim \omega_0 (g\_f - g\_i)^{x(g\_i)}$ with weakly $g\_i$ dependent exponent. We universally find $T\_f > T\_c(g\_f)$ [\cref{fig:quench_phase}], showing that no (thermal) superconductivity remains in the long time limit. 

Indeed, we generally find that it is not possible to induce superconductivity by interaction quenches in the YSYK model, neither transiently nor at late times. The dynamics of the superconducting order parameter scales as $\Delta(t) \sim \mathcal{O}(\alpha)$ for all times [\cref{fig:quench_gap}], so that for the physical limit $\alpha \to 0$, no superconducting fluctuations remain. Instead, the non-Fermi liquid normal state of the YSYK models rapidly heats following the interaction quench, prohibiting the formation of cooper pairs. Our surprising results remain true, when considering quenches with finite, linear interaction ramp \cite{supplementary}.

This is distinct from BCS theory, where one finds a coherently oscillating, yet finite order parameter following interaction quenches \cite{barankovCollectiveRabi2004}. Further, it deviates from observations made for similar protocols in Hubbard models, treated perturbatively or with DMFT, where a transiently ordered state can emerge following an interaction quench \cite{bauerDynamicalInstabilities2015, ecksteinInteractionQuench2010, stahlElectronicFluctuation2021, tsujiNonthermalAntiferromagnetic2013, wernerNonthermalSymmetrybroken2012}.

\paragraph*{\textbf{Dressed Statistics} ---}

\begin{figure}
    \centering
    \includegraphics[width = \columnwidth]{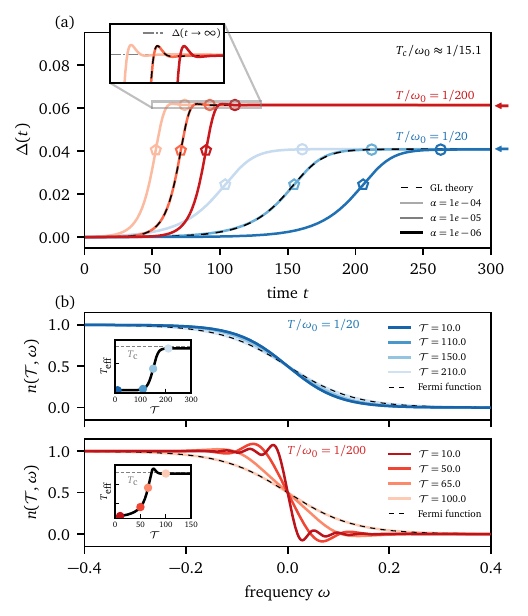}
    \caption{\textbf{Dynamical relaxation from undercooled normal state into superconductivity} at $g = 1$. (a) Order parameter $\Delta(t)$ [\cref{eq:ysyk_gap_function}] for different symmetry breaking strengths $\alpha$ [\cref{eq:symmetry_breaking}] and temperatures. Close to $T_c$ (blue) the dynamics is overdamped while for $T \ll T\_c$ (red) it becomes underdamped and oscillatory. Both regimes are well described by Ginzburg Landau theory (dashed lines). The color matched arrows mark the thermal $\Delta$ evaluated at $T\_{eff}(\infty)$, indicating thermalization. (b) Non-equilibrium occupation function $n(\mathcal{T}, \omega)$ [\cref{eq:ysyk_Teff}] for different center mass times $\mathcal{T}$ indicated as dots in the effective temperature inset. Close to $T\_c$ the system is locally thermal and follows a hydrodynamic time evolution. For $T \ll T\_c$ the early times are non-thermal and only at late times does the system thermalize.}
    \label{fig:undercooling}
    \label[fig_a]{fig:undercooling_gap}
    \label[fig_b]{fig:undercooling_occupation}
\end{figure}

Next, we consider the emergence of superconductivity due to light-induced cooling. This we model as the dynamical relaxation from an undercooled YSYK normal state, into the superconducting ground state, following an explicit U(1) symmetry breaking; see \cref{eq:symmetry_breaking}. An undercooled state, that is an unordered $\Delta = 0$ state with $T < T\_c$, is unstable to such perturbations and its exact relaxation dynamics is shown in \cref{fig:undercooling}. Similar to the quench protocol above, we find that the system universally thermalizes in the long time limit, but now with $T\_{eff}(\infty) < T_c$, leading to emergent superconducting states.

The exact dynamics of the superconducting order parameter in shown in \cref{fig:undercooling_gap}. While the thermalization time $t_*$ (circles) shows a strong dependence on $\alpha$, the overall dynamics as well as the final state $\Delta(t > t_*)$ are independent of the symmetry breaking strength \cite{supplementary}. The latter agrees with the thermal gap function evaluated for $T\_{eff}(t \to \infty)$ (color matched arrows), further illustrating thermalization at late times.

The early time dynamics ($t < \tilde{t}_*$; pentagon), is universally determined by an exponentially growing order parameter
\begin{equation}
    \Delta(t) \sim e^{\Gamma(T) t}
    \label{eq:thermalization_rate}
\end{equation}
with $\alpha$ independent realxation rate $\Gamma(T)$, while at later times ($t > \tilde{t}_*$)
the dynamics in the weak ($T \lesssim T\_c$) and strong ($T \ll T\_c$) undercooling regime is qualitatively different. Close to $T\_c$ it is underdamped, in the sense that $\delta(t) = \Delta(t) - \Delta(\infty)$ has no zero crossing, while for $T \ll T\_c$ we observe an oscillatory dynamics (\cref{fig:undercooling_gap}; inset). This difference can be understood intuitively. Close to $T\_c$, the superconducting and normal state do not deviate strongly
such that the `effective force', driving
the relaxation towards the ordered state is small compared to internal relaxation rates. Locally the system is in thermal equilibrium and the time evolution is hydrodynamic, with non-equilibrium occupation functions resembling Fermi-distributions (\cref{fig:undercooling_occupation}; upper panel). For $T \ll T\_c$ the `effective force' becomes large compared to the internal relaxation rate so that the system is non-thermal at short and intermediate times and only thermalizes in the long time limit (\cref{fig:undercooling_occupation}; lower panel).

The superconducting order parameter dynamics can be reproduced by time dependent Ginzburg-Landau theory (GL) with $F\_{GL} = c_1 \Delta^2 + c_2 \Delta^4$ and dynamical equation
\begin{align}
    \eta \partial_t^2 \Delta(t) + \delta \partial_t \Delta(t) = -\frac{\partial F\_{GL}}{\partial \Delta}.
\end{align}
For weak undercooling, the dynamics is described by overdamped GL ($\eta \to 0$), while the oscillatory case requires both finite $\eta, \delta$ to reproduce the exact data [dashed lines in \cref{fig:undercooling_gap}]. BCS theory corresponds to the limit $\delta \to 0$, implying that the interactions effectively introduce a damping for the order parameter dynamics \cite{supplementary}. 

\begin{figure}[t]
    \centering
    \includegraphics{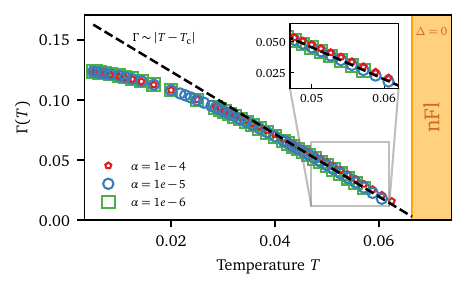}
    \caption{\textbf{Relaxation rate $\boldsymbol{\Gamma(T)}$} from the early time dynamics $\Delta(t) \sim e^{\Gamma(T) t}$ at $g = 1$. $\Gamma$ extracted at different symmetry breaking strengths $\alpha$ (different shapes) becomes indistinguishable for $\alpha < 10^{-5}$ illustrating the $\alpha$ independence. Close to $T\_c$ we find $\Gamma \sim |T - T\_c|$ in agreement with the overdamped Ginzburg-Landau prediction (dashed line). 
    Deep within the superconducting phase $T \ll T\_c$ the rate saturates.
    Note $\lim_{T \to T\_c} \Gamma(T) = 0$, indicating the critical slowing down.}
    \label{fig:relaxation_rate}
\end{figure}
The temperature dependence of the relaxation rate $\Gamma(T)$ is a defining characteristic of the relaxation process. For $T \lesssim T\_c$ it agrees well with the overdamped GL prediction $\Gamma(T) \sim |T - T\_c|$, while it saturates for $T \ll T\_c$ (\cref{fig:relaxation_rate}). The reason for the different regimes is again intuitively understood via the `effective force' argument. Close to $T\_c$ the speed of the relaxation dynamics is limited by the force, while for $T \ll T\_c$ internal relaxation processes determine the relaxation timescale. 

Generically, the relaxation rate vanishes as $\Gamma(T) \sim |T - T\_c|^\gamma$ with critical exponent $\gamma$, a phenomenon known as critical slowing down \cite{hohenbergTheoryDynamic1977,tinkhamIntroductionSuperconductivity2004}. From our exact dynamics we find $\gamma = 1$, which despite the strong correlations and nFl normal state, agrees with time dependent GL theory.


While the general structure of the relaxation is similar for different couplings $g$, we find that the dynamics is faster at strong coupling $\partial_g \Gamma(T) > 0$, but like the critical temperature saturates for $g \gtrsim 2$ \cite{esterlisCooperPairing2019,supplementary}. Further, the oscillatory order parameter dynamics is most pronounced for $g \lesssim 1$, while the amplitude of the oscillations becomes strongly suppressed at stronger interactions \cite{supplementary}.

\paragraph*{\textbf{Discussion} ---}
We studied the dynamical onset of superconductivity in a paradigmatic, solvable toy model for strongly correlated superconductivity by analyzing two dynamical protocols. These are motivated by the mechanism of light induced superconductivity; that is the dressing of Hamiltonian parameters and light-induced cooling.

Modeling the (quasi-static) dressing of Hamiltonian parameters, as interaction quenches across the superconducting phase transition we found that no superconductivity is induced due to excessive heating of the non-Fermi liquid normal state. This contrasts results obtained for BCS theory and DMFT studies in the Hubbard model.

Imitating light-induced cooling by analyzing the dynamical relaxation from an undercooled YSYK normal state, we universally observed superconductivity at late times. Depending on the strength of the undercooling, we found an overdamped or oscillatory relaxation dynamics towards these states. The resulting order parameter dynamics could be reproduced by time dependent Ginzburg Landau theory. 

An exciting extension of our work is the investigation of periodically and parametrically driven YSYK models, as this protocol has been suggested to be able to induce superconductivity \cite{babadiTheoryParametrically2017, eckhardtTheoryResonantly2023, knapDynamicalCooper2016,murakamiNonequilibriumSteady2017}. Our studies of the driven YSYK model revealed that driving leads to excessive heating, prohibiting the formation of superconductivity, but the additional coupling to a thermal bath \cite{zhangEvaporationDynamics2019,cheipeshQuantumTunneling2021,larzulAreFast2022, larzulEnergyTransport2022, larzulQuenchesPre2022} might mitigate this effect and lead to Floquet steady-states \cite{iadecolaFloquetSystems2015, langemeyerEnergyFlow2014} that could host non-thermal light-induced superconductivity. This promises to further enhance our understanding of the non-equilibrium superconductivity in strongly correlated systems.




\paragraph*{Note Added:} After the completion of this work we learned about an independent study of the non-equilibrium dynamics of the YSYK normal state, following a quench coupling to an external reservoir \cite{hosseinabadiThermalizationNonFermi2023}.

\paragraph*{\textbf{Acknowledgements} ---}
We thank V. Meden, E. Esterlis, C. Kuhlenkamp, M. Eckstein, D. Valentinis, E. V. Bostroem and A. Rubio for insightful discussions. Simulations were performed with computing resources granted by RWTH Aachen University under projects thes0823, rwth1408, and on the HPC system Raven at the Max Planck Computing and Data Facility. This work was supported by the Deutsche Forschungsgemeinschaft (DFG,
German Research Foundation) under RTG 1995, within the Priority Program
SPP 2244 ``2DMP'' --- 443273985 and under Germany's Excellence Strategy - Cluster of
Excellence Matter and Light for Quantum Computing (ML4Q) EXC 2004/1 -
390534769. We also acknowledge support by the Max Planck-New York City Center for Non-Equilibrium Quantum Phenomena.

\nocite{
    zinn-justinQuantumField2002,
    kamenevFieldTheory2011,
    rammerQuantumField2007,
    langrethTheorySpin1972,
    presswilliamh.ComputingAccurate1989,  
    songStronglyCorrelated2017a,
    fornbergImprovingAccuracy2021,
    stefanucciNonequilibriumManyBody2013,
    butcherNumericalMethods2016,
    hairerSolvingOrdinary2000,
    mogensenOptimMathematical2018,
    rackauckasDifferentialequationsJl2017}  

\bibliographystyle{apsrev4-2}
\bibliography{references_syk.bib}

\end{document}


\title{Supplementary Material for:\\ Dynamical Onset of Light-Induced Unconventional Superconductivity -- \\ a Yukawa-Sachdev-Ye-Kitaev study}

\author{Lukas Grunwald}
\email{lukas.grunwald@mpsd.mpg.de}
\affiliation{\affiliationRWTH}
\affiliation{\affiliationMPSD}

\author{Giacomo Passetti}
\affiliation{\affiliationRWTH}

\author{Dante M.~Kennes}
\email{dante.kennes@mpsd.mpg.de}
\affiliation{\affiliationRWTH}
\affiliation{\affiliationMPSD}

\date{\today}

\maketitle
\onecolumngrid
\tableofcontents

\section{Exact Solution of YSYK on the Keldysh Contour}
We derive the exact solution of the YSYK model on the Keldysh contour at large $N$. For completeness, we consider a slight generalization to the model introduced in the main text by considering spin non-diagonal interactions as well as a finite chemical potential. The Hamiltonian reads \cite{esterlisCooperPairing2019,wangSolvableStrongCoupling2020,choiPairingInstabilities2022}
\begin{equation}
    H = -\mu \sum_{i=1}^N c_{i\alpha}^\dagger c_{i \alpha} +
    \frac{1}{2}\sum_{k=1}^N \big( \pi_k^2 + \omega_0^2 \phi_k^2 \big)
    + \frac{1}{N}\sum_{ij} \sum_k f(t) \tilde{g}_{ijk} \phi_k \, 
    c_{i\alpha}^\dagger (\sigma\^x_{\alpha \beta})c_{j\beta},
    \label{eq:ysyk_hamiltonian}
\end{equation}
where $\sigma_{\alpha \beta}\^x, x = 0,\dots, 3$ denotes the Pauli-matrices and $f(t)$ a generic time dependent function. We will derive the general solution of \cref{eq:ysyk_hamiltonian} and restrict to our specific model with $\textnormal{x}, \mu =0$ in the end. The starting point is the path-integral on the Keldysh-contour \cite{kamenevFieldTheory2011,rammerQuantumField2007}. For the YSYK model, the associated action appearing in the generating functional $Z = \int \mathcal{D}[\bar{\psi},\psi] \mathcal{D}[\phi] \exp{iS[\psi, \phi]}$ reads
\begin{align}
    &S[\psi, \phi] = S_0[\psi, \phi] + S\_{int}[\psi, \phi], \\
    &S_0[\psi, \phi] = \int_\mathcal{C} \Biggl\{  \sum_i \bar{\psi}_{i \alpha}(t)
    \big( i \partial_t + \mu \big) \psi_\alpha(t)
    + \frac{1}{2} \sum_k \phi_k(t) \bigl( -\partial_t^2 - \omega_0^2 \bigr) \phi_k(t)
    \Biggr\}, \\
    &S\_{int}[\psi, \phi] =-\frac{1}{N} \int_\mathcal{C} \sum_{ijk} f(t) \tilde{g}_{ij, k} \phi_k(t) \bar{\psi}_{i\alpha}(t) (\sigma\^x_{\alpha \beta}) \psi_{j\beta}(t).
\end{align}
We imply a summation over repeated Greek indices and use the shorthand $\int_{\mathcal{C}^{(')}} (\dots) = \int_\mathcal{C} \dd t^{(')} (\dots)$ for integrals over the Keldysh-contour $\mathcal{C} = \mathcal{C}_+ \cup \mathcal{C}_-$, which consists  of forward branch $\mathcal{C}_+$, running from $t_0 \to \infty$ and backward branch $\mathcal{C}_-$ running back from $\infty \to t_0$. We assume no initial correlations and set $t_0 \to -\infty$ \cite{rammerQuantumField2007,kamenevFieldTheory2011}, which is sufficient for the non-equilibrium protocols we analyze in this work (see below). 

From the non-interacting action $S_0$, we can directly extract the free fermionic and bosonic contour propagators
\begin{align}
    (G_0^{-1})_{\alpha \beta}(t,t') = \bigl[ i \partial_t + \mu \bigr] \delta(t,t') \delta_{\alpha \beta}, \qquad
    D_0^{-1}(t,t') = \bigl[ -\partial^2_t - \omega_0^2 \bigr] \delta(t,t'),
\end{align}
with $\int_\mathcal{C'} \delta(t, t') = 1$. Next we execute the disorder average over the interaction vertices $g_{ij,k} \in $ GOE (Gaussian-Orthogonal-Ensemble \cite{akemannOxfordHandbook2015}), where for each bosonic index $k$, $g_{ij,k}$ represents a random matrix in the indices $i,j$. The GOE is an ensemble of symmetric random matrices $g_{ij}$ (suppress bosonic index k), which is defined over the mean $\overline{g_{ij}} = 0$ and the variance $\overline{g_{ij} g_{km}} = (\delta_{ik}\delta_{jm} + \delta_{im}\delta_{jk})g^2$ of the matrices. The probability density and measure read
\begin{align}
    \rho(g) \sim \text{exp} \left(- \frac{1}{4g^2} \Tr(g^2) \right), \qquad
    d\mu(g) \sim \prod_{i \leq j} d(g_{ij}),
\end{align}
so that the evaluation of the disorder average reduces to calculating Gaussian integrals (no replica trick is needed in the Keldysh formalism \cite{kamenevFieldTheory2011}). The resulting disorder averaged action reads
\begin{align}
    \label{eq:ysyk_disorder_avg_action}
    iS\_{avg}[\psi, \phi] = &\int_\mathcal{C} 
    \Biggl\{  \sum_i \bar{\psi}_{i \alpha}(t)
    \big( i \partial_t + \mu \big) \psi_\alpha(t)
    + \frac{1}{2} \sum_k \phi_k(t) \bigl( -\partial_t^2 - \omega_0^2 \bigr) \phi_k(t)
    \Biggr\} \\ \nonumber
    +
    \frac{g^2}{2N^2} \int_\mathcal{C} &\int_{\mathcal{C}'} 
        f(t) f(t') \sum_k \phi_k(t) \phi_k(t')
        \left\{
            \bigg[\sum_i \big(\psi_{i, \beta} \bar{\psi}_{i, \alpha'} \big)(t,t') \bigg]
            \left( \sigma\^x_{\alpha', \beta'} \right)
            \bigg[\sum_i \big(\psi_{j, \beta'} \bar{\psi}_{j, \alpha} \big)(t',t) \bigg]
            \left( \sigma\^x_{\alpha, \beta} \right) \right.
            \\ \nonumber
            & \hspace{4.05cm}-
            \left.
             \bigg[\sum_i \big(\bar{\psi}_{i, \alpha} \bar{\psi}_{i, \alpha'} \big)(t,t')\bigg]
             \left( \sigma\^x_{\alpha', \beta'} \right)
             \bigg[\sum_j \big(\psi_{i, \beta'} \psi_{i, \beta} \big)(t',t)\bigg]
             \left( \sigma\^x_{\alpha, \beta} \right)
     \right\}.
\end{align}
Note the appearance of anomalous terms $\sim \psi_\alpha(t) \psi_\beta(t')$, which give rise to superconductivity and which are entirely due to $g_{ij,k} \in$ GOE being real. For complex $g_{ij,k} \in$ GUE (Gaussian-Unitary-Ensemble), like in the SYK model \cite{chowdhurySachdevYeKitaevModels2022}, we would obtain \cref{eq:ysyk_disorder_avg_action}, but without the anomalous contributions. Next, we introduce the site averaged contour Greens functions and corresponding self-energies (Lagrange-multipliers) as dynamical mean-fields via fat-unities
\begin{align}
    \label{eq:ysyk_fat_unity_1}
    &1 = \int \mathcal{D}[\Sigma, G] \exp \Biggl( 
        N \int_\mathcal{C} \int_{\mathcal{C}'} \Sigma_{\alpha \beta}(t, t') 
        \biggl\{G + \frac{i}{N}\sum_i \psi_i \bar{\psi}_i  \biggr\}_{\beta \alpha}(t', t)
    \Biggr), \\
    \label{eq:ysyk_fat_unity_2}
    &1 = \int \mathcal{D}[\Pi, D] \exp \Biggl( -\frac{N}{2}
        \int_\mathcal{C} \int_{\mathcal{C}'} \Pi(t, t') 
        \biggl\{D + \frac{i}{N}\sum_k \phi_k \phi_k  \biggr\}(t', t)
    \Biggr), \\
    \label{eq:ysyk_fat_unity_3}
    &1 = \int \mathcal{D}[\bar{\phi}, F] \exp \Biggl( \frac{N}{2} 
        \int_\mathcal{C} \int_{\mathcal{C}'} \bar{\phi}_{\alpha \beta}(t, t') 
        \biggl\{F + \frac{i}{N}\sum_i \psi_i \psi_i  \biggr\}_{\beta \alpha}(t', t)
    \Biggr), \\
    \label{eq:ysyk_fat_unity_4}
    &1 = \int \mathcal{D}[\phi, \bar{F}] \exp \Biggl( \frac{N}{2} 
    \int_\mathcal{C} \int_{\mathcal{C}'} \phi_{\alpha \beta}(t, t') 
    \biggl\{\bar{F} + \frac{i}{N}\sum_i \bar{\psi}_i \bar{\psi}_i  \biggr\}_{\beta \alpha}(t', t)
    \Biggr).
\end{align}
The signs and prefactors are chosen in anticipation of obtaining Dyson equations (see below). Using the dynamical mean-fields, one finds for the interaction part ($\sim g^2$) of the averaged action
\begin{equation}
    \frac{iS\_{avg,int}}{N} = - \frac{ig^2}{2} \int_\mathcal{C} \int_{\mathcal{C}'} f(t) f(t') D(t,t') \tr \Bigl[
        G_{\cdot, \cdot}(t,t') \sigma\^x G_{\cdot, \cdot}(t', t) \sigma\^x - \bar{F}_{\cdot, \cdot}(t,t') \sigma\^x F_{\cdot, \cdot}(t', t) (\sigma\^x)\^T \Bigr],
\end{equation}
where the $\tr[\dots]$ is to be taken over the spin indices. The $\psi,\bar{\psi}$ and $\phi$ fields are now decoupled and can be integrated out separately after taking into account the self-energy contributions from the fat unities. This process is straight forward in the bosonic case
\begin{equation}
    \int \mathcal{D}[\phi] \exp \Biggl\{ i\int_\mathcal{C} \int_{\mathcal{C}'} \sum_k \phi_k(t') \bigl[ D_0^{-1}(t,t') - \Pi(t,t')\bigr] \phi_k(t') \Biggr\} \sim \exp \biggl\{-\frac{N}{2} \log \det \bigl[ D_0^{-1} - \Pi \bigr]  \biggr\}.
\end{equation}
In the fermionic sector, the action contains both normal $\sim \bar{\psi}_\alpha \psi_\beta$ and anomal $\sim \psi_\alpha \psi_\beta$ contributions, the latter stemming from the fat unities \cref{eq:ysyk_fat_unity_3,eq:ysyk_fat_unity_4}. We introduce Nambu vectors $\vec{\psi}_i = (\psi_{i,\uparrow}, \psi_{i,\downarrow}, \bar{\psi}_{i,\uparrow}, \bar{\psi}_{i, \downarrow})$ to rewrite the remaining fermionic action as a Grassman-Gaussian integral
\begin{gather}
    \label{eq:ysyk_action_free_fermions}
    \int \mathcal{D} [\vec{\psi}] \; \exp
    \left\{
    \frac{i}{2}\int_\mathcal{C} \int_{\mathcal{C}'} \sum_i
    \vec{\psi}_i^\dagger(t) \bigl[ \tilde{\boldsymbol{G}}_0^{-1}(t,t') - \tilde{\boldsymbol{\Sigma}}(t,t') \bigr] \vec{\psi}(t')
    \right\} \\[1pt]
    \textnormal{with} \quad 
    \tilde{\boldsymbol{G}}_0^{-1}(t, t') = \mqty(
        (G_0^{-1})_{\cdot,\cdot} & 0 \\
        0 & -(G_0^{-T})_{\cdot,\cdot})(t,t'),
    \qquad
    \tilde{\boldsymbol{\Sigma}}(t, t') = \mqty(
    \Sigma_{\cdot,\cdot} & \phi_{\cdot,\cdot} \\
    \bar{\phi}_{\cdot,\cdot} & -\Sigma_{\cdot,\cdot}\^T )(t, t').
\end{gather}
Here we introduced the Nambu-Keldysh self-energy $\tilde{\boldsymbol{\Sigma}}(t,t')$ and free-propagator $\tilde{\boldsymbol{G}}_0(t,t')$ and understand the transpose as $A_{\alpha \beta}(t,t')\^T = A_{\beta \alpha}(t',t)$. Note that the integral-measure only contains $\vec{\psi}$ but not $\vec{\psi}^\dagger$, since the latter is just a rearrangement of the former. Hence, \cref{eq:ysyk_action_free_fermions} is not an ordinary Grassman-Gaussian integral. For $M \in \textnormal{Skew}_{n}$ and Grassman fields $\vec{\theta} = (\theta_1, \dots, \theta_{n})$ integrals of this type equate to \cite{zinn-justinQuantumField2002}
\begin{equation}
    \int \dd \vec{\theta} \exp(-\frac{1}{2} \vec{\theta}\^T M \vec{\theta}) =
    \begin{cases}
        \textnormal{Pf}(M) & n \text{ even}\\
        0 & n \text{ odd} \; .
    \end{cases}
    \label{eq:ysyk_pfaffian_integral}
\end{equation}
\Cref{eq:ysyk_action_free_fermions} can be brought into exactly this form by permuting elements in $[\dots]$, and we always have even $n$ because of the 4-component Nambu vectors. Using \cref{eq:ysyk_pfaffian_integral} and reversing the permutation in the $\log\textnormal{Pf}[\dots]$ term, we find
\begin{equation}
    \int \mathcal{D} [\vec{\psi}] \; \exp
    \left\{
    \frac{i}{2}\int_\mathcal{C} \int_{\mathcal{C}'} \sum_i
    \vec{\psi}_i^\dagger(t) \bigl[ \tilde{\boldsymbol{G}}_0^{-1}(t,t') - \tilde{\boldsymbol{\Sigma}}(t,t') \bigr] \vec{\psi}(t')
    \right\}
    \sim \exp 
    \Bigl\{
        N \log \textnormal{Pf} \, \bigl[ \tilde{\boldsymbol{G}}_0^{-1} - \tilde{\boldsymbol{\Sigma}} \bigr]
    \Bigr\}.
\end{equation}
The resulting $G\Sigma$-action only depends on the dynamical mean fields $G,\Sigma,F, \phi, \bar{F}, \bar{\phi}$ and reads
\begin{align}
    \frac{iS_{G\Sigma}}{N} =  
    &\log \textnormal{Pf} \, \bigl[ \tilde{\boldsymbol{G}}_0^{-1} - \tilde{\boldsymbol{\Sigma}} \bigr]
    -\frac{1}{2} \log \det \bigl[ D_0^{-1} - \Pi \bigr]
    + \frac{1}{2} \Tr\bigl( \tilde{\boldsymbol{\Sigma}} \star \tilde{\boldsymbol{G}} \bigr) - \frac{1}{2} \Tr\bigl( \Pi \star D \bigr) \nonumber \\
    &- \frac{ig^2}{2} \int_\mathcal{C} \int_{\mathcal{C}'} f(t) f(t') D(t,t') \tr \Bigl[
        G_{\cdot, \cdot}(t,t') \sigma\^x G_{\cdot, \cdot}(t', t) \sigma\^x - \bar{F}_{\cdot, \cdot}(t,t') \sigma\^x F_{\cdot, \cdot}(t', t) (\sigma\^x)\^T \Bigr],
    \label{eq:ysyk_gsigma_action}
\end{align}
where the $\Tr(\dots)$ is to be taken over all indices and times and where we arranged fermionic propagators $G,F,\bar{F}$ into the Nambu-Keldysh Greens function
\begin{equation}
    \tilde{\boldsymbol{G}}(t,t') = \mqty( 
        G_{\cdot,\cdot} & F_{\cdot, \cdot}\\
        \bar{F}_{\cdot, \cdot} & -G_{\cdot, \cdot}^{\textnormal{T}}
    )(t,t').
\end{equation}
The shorthand $(. \star .)$ abbreviates  convolutions over the support of the functions involved, in this case the Keldysh-contour $\mathcal{C}$ as well as the matrix structure of the GF's.

The action is globally multiplied by $N$, such that in the large $N$ limit, the exact solution can be obtained by evaluating the variation $\delta S_{G\Sigma}$. This yields equations of motion for the dynamical mean-fields. Similar to derivatives of $\log \det (\dots)$ terms, (functional) derivatives of Pfaffians can be calculated as \cite{zinn-justinQuantumField2002}
\begin{equation}
    \frac{1}{\textnormal{Pf}(A)} \frac{\partial \textnormal{Pf}(A)}{\partial A_{ij}} = \frac{1}{2} \tr \left[ A^{-1} \frac{\partial A}{ \partial A_{ij}} \right].
\end{equation}
Varying the action with respect to the self-energies $\Sigma, \phi, \bar{\phi}$ and $\Pi$, we obtain the contour Dyson equations \cite{kamenevFieldTheory2011,rammerQuantumField2007} for the Numbu-Keldysh GF's and the bosonic propagator respectively
\begin{align}
    \tilde{\boldsymbol{G}}(t,t') = \big( \tilde{\boldsymbol{G}}_0^{-1} - \tilde{\boldsymbol{\Sigma}} \big)^{-1}(t,t'),
    \qquad
    D(t,t') = \big(D_0^{-1} - \Pi \big)^{-1}(t,t').
    \label{eq:ysyk_sde_dyson}
\end{align}
During the evaluation of the former we used $[ \tilde{\boldsymbol{G}}_0^{-1} - \tilde{\boldsymbol{\Sigma}}]_{11}^{-1} = -[ \tilde{\boldsymbol{G}}_0^{-1} - \tilde{\boldsymbol{\Sigma}}]_{22}^{-\textnormal{T}}$. The equations for the self-energies follow by varying $S_{G\Sigma}$ with respect to the Greens functions $G, F, \bar{F}, D$, resulting in
\begin{align}
    \label{eq:ysyk_sde_full_1}
    &\Sigma_{\alpha \beta}(t,t') = i g^2 f(t)f(t') D(t, t')
     \bigl[\sigma\^x G_{\cdot,\cdot}(t,t') \sigma\^x \bigr]_{\alpha \beta},  \\
     \label{eq:ysyk_sde_full_2}
    &\phi_{\alpha \beta}(t,t')  = -ig^2 f(t)f(t') D(t, t') \bigl[\sigma\^x F_{\cdot,\cdot}(t,t') (\sigma\^x)\^T \bigr]_{\alpha \beta},  \\
    \label{eq:ysyk_sde_full_3}
    &\bar{\phi}_{\alpha \beta}(t,t')  = -ig^2 f(t)f(t') D(t, t') \bigl[(\sigma\^x)\^T  \bar{F}_{\cdot,\cdot}(t,t')  \sigma\^x \bigr]_{\alpha \beta}, \\
    \label{eq:ysyk_sde_full_4}
    &\Pi(t, t') = -ig^2 f(t)f(t') \tr \Bigl[
        G_{\cdot,\cdot}(t,t') \sigma\^x G_{\cdot,\cdot}(t', t) \sigma\^x - \bar{F}_{\cdot,\cdot}(t,t') \sigma\^x F_{\cdot,\cdot}(t', t) (\sigma\^x)\^T \Bigr].
\end{align}
\Crefrange{eq:ysyk_sde_full_1}{eq:ysyk_sde_full_4} determine the exact self-energies of the YSYK model on the Keldysh contour ($t, t' \in \mathcal{C}$). We additionally obtained them using a diagrammatic derivation as a consistency check. Note that this system of equations only depends on the site averaged GF's and self-energies. This illustrates the self-averaging property, since the exact solution has no reference to the site indices of the original theory.

The hence derived equations contain both singlet and triplet pairing components. As in previous works on the YSYK model, we focus on the $\textnormal{x} = 0$ case, where the model has a $\textnormal{SU}(2)$ symmetry. The presence of triplet superconductivity would spontaneously break this symmetry. Following \cite{esterlisCooperPairing2019,wangSolvableStrongCoupling2020}, we assume no spontaneous $\textnormal{SU}(2)$ symmetry breaking and will thus focus on singlet anomalous propagators. On the Keldysh contour this implies $F_{\cdot ,\cdot} = F\_s i \sigma_y$ with $F\_s = \frac{F_{\uparrow \downarrow} - F_{\downarrow \uparrow}}{2}$. Further, due to the $\textnormal{SU}(2)$ symmetry, the normal GF and self-energy become spin independent, i.e. $G_{\cdot, \cdot}(t,t') = G(t,t') \sigma_0$ and $\Sigma_{\cdot,\cdot}(t,t') = \Sigma (t,t')\sigma_0$. 
Henceforth, we can work with reduced Nambu-vectors $\psi_i = (c_{i\uparrow}, c^\dagger_{i \downarrow})\^T$ and the associated contour GF's and self-energies ($t,t' \in \mathcal{C}$)
\begin{align}
    \boldsymbol{G}(t, t')
    &= -\frac{i}{N} \sum_i \overline{\expval*{T_\mathcal{C} \{ \psi_i(t)\psi^\dagger_i(t') \}}}
    = \mqty( 
        G(t,t') & F\_s(t,t')\\
        -\bar{F}\_s(t,t') & -G(t',t)
    ) \\ \label{eq:supp_contour_gf}
    \boldsymbol{\Sigma}(t, t') 
    &= \mqty( 
        \Sigma(t,t') & \phi\_s(t,t')\\
        -\bar{\phi}\_s(t,t') & -\Sigma(t',t)).
\end{align}
\Crefrange{eq:ysyk_sde_full_1}{eq:ysyk_sde_full_4} then simplify to the equations from the main text ($t, t' \in \mathcal{C}$)
\begin{align}
    \label{eq:supp_sde1}
    \boldsymbol{G}(t,t') &= \big( \boldsymbol{G}_0^{-1} - \boldsymbol{\Sigma} \big)^{-1}(t,t'), \\ \label{eq:supp_sde2}
    D(t,t') &= \big(D_0^{-1} - \Pi \big)^{-1}(t,t') \\ \label{eq:supp_sde3}
    \boldsymbol{\Sigma}(t,t') &= i g^2 f(t) f(t') D(t,t')
    \left[\sigma_3 \boldsymbol{G}(t,t') \sigma_3 \right], \\ \label{eq:supp_sde4}
    \Pi(t,t') &= -ig^2 f(t) f(t') \tr 
    \left[\sigma_3 \boldsymbol{G}(t,t') \sigma_3 \boldsymbol{G}(t',t) \right],
\end{align}
where $ \boldsymbol{G}_0^{-1}(t,t') = \left[i\partial_t + \mu \boldsymbol{\sigma}_3 \right] \delta(t,t')$. In order to solve \crefrange{eq:supp_sde1}{eq:supp_sde4}, we need to analytically continue them to real-times, and physical Greens functions $\boldsymbol{G}^\gtrless, D^\gtrless$ or $\boldsymbol{G}\^{R/A/K}, D\^{R/A/K}$. The latter are obtained by fixing the position of the contour times $t,t' \in \mathcal{C}$ on the Keldysh contour $t,t' \in \mathcal{C_\pm}$, viz. (similar for $D^\gtrless$) \cite{rammerQuantumField2007}
\begin{align}
    \boldsymbol{G}^>(t,t') &= \boldsymbol{G}(t,t'), \quad \text{with} \quad
    t \in \mathcal{C}_-, t' \in \mathcal{C}_{+}, \\
    \boldsymbol{G}^<(t,t') &= \boldsymbol{G}(t,t'), \quad \text{with} \quad
    t \in \mathcal{C}_+, t' \in \mathcal{C}_{-}, \\
    \boldsymbol{G}\^{R}(t, t') &= \theta(t-t') 
    \left( \boldsymbol{G}^> - \boldsymbol{G}^< \right)(t,t'), \\   
    \boldsymbol{G}\^{A}(t, t') &= \theta(t'-t) 
    \big( \boldsymbol{G}^< - \boldsymbol{G}^> \big)(t,t'), \\   
    \boldsymbol{G}\^{K}(t, t') &= 
    \big( \boldsymbol{G}^< + \boldsymbol{G}^> \big)(t,t').
\end{align}
Using the Langreth theorems \cite{langrethTheorySpin1972,rammerQuantumField2007}, \crefrange{eq:supp_sde1}{eq:supp_sde4} translate to a system of non-linear Volterra integro-differential equations for $\boldsymbol{G}^\gtrless$ and $D^\gtrless$ --- the so called Kadanoff-Baym equations (KBe's)
\begin{align}
    \label{eq:supp_final1}
    \bigl(
        i\partial_t + \mu \boldsymbol{\sigma}_3
    \bigr)
    \boldsymbol{G}^{\gtrless}(t, t')
    &= 
    \Bigl(
         \boldsymbol{\Sigma}\^R \star \boldsymbol{G}^{\gtrless} + 
         \boldsymbol{\Sigma}^{\gtrless} \star \boldsymbol{G}\^A 
    \Bigr)(t,t')
    \equiv \boldsymbol{\mathcal{F}}^{\gtrless}_1(t,t'), 
    \\ \label{eq:supp_final2}
    \boldsymbol{G}^{\gtrless}(t, t') 
    \bigl(
        -i \cev \partial _{t'} + \mu \boldsymbol{\sigma}_3
    \bigr) 
    &=
    \Bigl( 
        \boldsymbol{G}\^R \star \boldsymbol{\Sigma}^{\gtrless} + \boldsymbol{G}^{\gtrless} \star \boldsymbol{\Sigma}\^A 
    \Bigr)(t,t') \equiv \boldsymbol{\mathcal{F}}^{\gtrless}_2(t,t'), 
    \\ \label{eq:supp_final3}
    \Bigl(
         -\partial_t^2 - \omega_0^2 
    \Bigr) 
    D^{\gtrless}(t,t') 
    &=
    \Bigl(
         \Pi\^R \star  D^{\gtrless} + \Pi^{\gtrless} \star  D\^A
    \Bigr)(t,t') \equiv \mathcal{B}^{\gtrless}_1(t,t'),
    \\ \label{eq:supp_final4}
    D^{\gtrless}(t, t')
    \Bigl(
         - \cev \partial^2 _{t'} - \omega_0^2
    \Bigr) 
    &=
    \left(
         D\^R \star \Pi^{\gtrless} + D^{\gtrless} \star \Pi\^A
    \right)(t,t') \equiv \mathcal{B}^{\gtrless}_2(t,t'), 
    \\ \label{eq:supp_final5}
    \boldsymbol{\Sigma}^{\gtrless}(t,t') &= i g^2 f(t) f(t') D^\gtrless(t,t')
    \left[\sigma_3 \boldsymbol{G}^\gtrless(t,t') \sigma_3 \right], \\ \label{eq:supp_final6}
    \Pi^\gtrless(t,t') &= -ig^2 f(t) f(t') \tr 
    \left[\sigma_3 \boldsymbol{G}^\gtrless(t,t') \sigma_3 \boldsymbol{G}^\lessgtr(t',t) \right],
\end{align}
where $(A \star B)(t,t') = \int_{t_0}^\infty A(t, \tau) B(\tau, t') \dd{\tau}$ and matrix multiplications are implied. Due to the causality structure of the GF's, these equations can be integrated as an initial value problem (see next section).

In this work the initial conditions are fixed in the lower quadrant of the two-time plane  $t, t' < 0$ as the equilibrium solution of \crefrange{eq:supp_final1}{eq:supp_final6}, i.e. for $f(t) \equiv \textnormal{const}$. In equilibrium the GF's only depend on time differences, allowing us to map the differential equations' [\crefrange{eq:supp_final1}{eq:supp_final4}] onto algebraic equations in frequency space. There self-energies and Greens functions are connected by
\begin{align}
    \boldsymbol{G}\^R(\omega) = \frac{1}{(\omega + i0^+) + \mu \boldsymbol{\sigma}_3 - \boldsymbol{\Sigma}\^R(\omega)}, \qquad
    D\^R(\omega) = \frac{1}{(\omega + i0^+)^2 - \omega_0^2 - \Pi\^R(\omega)}.
    \label{eq:supp_dyson}
\end{align}
Further $\boldsymbol{G}\^R(\omega), \boldsymbol{G}\^K(\omega)$ and $D\^R, D\^K$ are related by fluctuation dissipation theorems respectively \cite{rammerQuantumField2007}. Hence, \cref{eq:supp_dyson} together with \cref{eq:supp_final4,eq:supp_final5} provides a closed, algebraic system of equations for the equilibrium GF's. It can be solved numerically using a fixed point iteration in combination with numerical Fourier transforms, which we evaluate using an interpolation based FFT-scheme \cite{presswilliamh.ComputingAccurate1989}. The fixed point iteration is analog to the SYK model and previous studies of the YSYK model; see \cite{chowdhurySachdevYeKitaevModels2022,songStronglyCorrelated2017a,esterlisCooperPairing2019,valentinisCorrelationPhase2023} and references therein.  

\section{Integration of Kadanoff-Baym equations}
\paragraph*{\textbf{General Idea} ---}
In this section we discuss the general procedure -- time stepping scheme -- for solving Kadanoff-Baym equations. Since the general time-stepping is the same for bosons and fermions, we explain our scheme using the generic KBe's
\begin{align}
    \label{eq:kbe1_down}
    \bigl[ i \partial_t - h(t) \bigr] G^{\gtrless}(t, t') &=
    \bigl( \Sigma\^R \star G^{\gtrless} + \Sigma^{\gtrless} \star G\^A \bigr)(t,t')
    \equiv \mathcal{F}^{\gtrless}_1(t,t'), \\
    G^{\gtrless}(t, t') \bigl[-i \cev \partial _{t'} - h(t')\bigr] &= 
    \bigl( G\^R \star \Sigma^{\gtrless} + G^{\gtrless} \star \Sigma\^A \bigr)(t, t')
    \equiv \mathcal{F}^{\gtrless}_2(t,t'),
    \label{eq:kbe1_right}
\end{align}
where $h(t)$ denotes the non-interacting Hamiltonian and the matrix structure is left implicit. The bosonic equations encountered in the last section are equivalent, apart from $\partial_t \to \partial_t^2$. We will comment on the implications of this below.

\begin{SCfigure}[][b]
    \centering
    \includegraphics{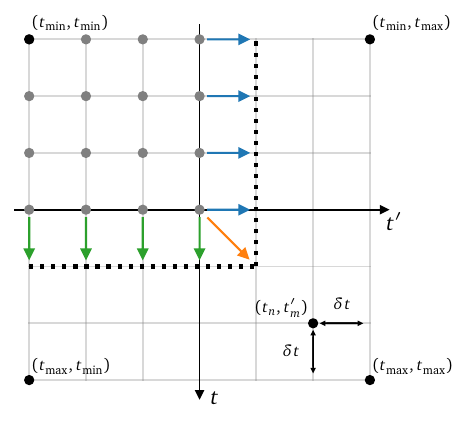}    
    \caption{\textbf{KBe time stepping scheme}. Given $G^>(t,t'), G^<(t,t') $ on an initial interval $t,t' \in (t_0, t_1)$, here illustrated as gray dots, we can obtain the solution at all times by time-stepping with the KBe's. One time step comprises the calculation of the points indicated by the 
    green $G^{\gtrless}(t_1 + \delta t, t' \in (t_0, t_1))$,
    blue  $G^{\gtrless}(t \in (t_0, t_1), t_1 + \delta t)$, and
    orange $G^{\gtrless}(t_1 + \delta t, t_1 + \delta t)$ arrows, corresponding to the propagation with \eqref{eq:kbe1_down}, \eqref{eq:kbe1_right}, \eqref{eq:kbe1_diag} respectively.}
    \label{fig:KBe_stepping}
\end{SCfigure}

Because of the causality structure of $G\^{R/A}$ and $\Sigma\^{R/A}$, the r.h.s of the KBe's for $G^{\gtrless}(t, t')$ only depends on times smaller than $\max(t,t')$. Thus, providing initial conditions for $G^{\gtrless}(t, t')$ in some interval $t, t' \in (t_0, t_1)$, the KBe's can be integrated as an initial value problem \cite{stefanucciNonequilibriumManyBody2013}. Due to them being integro-differential equations, each time-step depends on \textit{all} previous times via the history integrals $\mathcal{F}^{\gtrless}_1, \mathcal{F}^{\gtrless}_2$. A time-step is defined as advancing $\delta t$ forward in time, naturally enforcing a time discretization  $t_n = t_0 + n \delta t$. The different KBe's correspond to different directions in the $(t,t')$-plane and in order to perform a time-step, they need to be used in combination. This we illustrate in \cref{fig:KBe_stepping}. \Cref{eq:kbe1_down} propagates the Greens functions \textit{down} [green, $G^{\gtrless}(t_1 + \delta t, t' \in (t_0, t_1))$] and \cref{eq:kbe1_right} propagates to the \textit{right} [blue, $G^{\gtrless}(t \in (t_0, t_1), t_1 + \delta t)$]. For the diagonal propagation [orange, $G^{\gtrless}(t_1 + \delta t, t_1 + \delta t)$] we subtract \cref{eq:kbe1_right,eq:kbe1_down} to obtain
\begin{equation}
    i \partial_t G^{\gtrless}(t, t) =
    \bigl[ h(t), G^{\gtrless}(t, t) \bigr] +
    \bigl( \mathcal{F}^{\gtrless}_1 - \mathcal{F}^{\gtrless}_2 \bigr)(t,t').
    \label{eq:kbe1_diag}
\end{equation}

\Cref{eq:kbe1_down,eq:kbe1_right,eq:kbe1_diag} allow the calculation of all points needed for further time steps (dotted line in \cref{fig:KBe_stepping}). They can thus be used  to propagate the GF's in the entire $(t,t')$-plane, given some initial conditions. The full integration of the KBe's yields $G^{\gtrless}(t, t')$ for $t, t' \in (t_0, t\_{max})$.

The bosonic equations [\cref{eq:supp_final3,eq:supp_final4}] have exactly the same analytical structure as \cref{eq:kbe1_down,eq:kbe1_right}, and we can hence use the same scheme as outlined above with one caveat. Because of $\partial_t^2$ it is not possible to generate an equation for the diagonal, comparable to \cref{eq:kbe1_diag}, which only contains a second order derivative. Instead, the diagonal is determined as a combination of propagation to the \textit{right} and \textit{down} [\cref{eq:supp_final3,eq:supp_final4}].

\paragraph*{\textbf{Implementation} ---} We solve the KBe's on the interval $[t_0, t\_{max}]$. Formally we have $t_0 \to - \infty$, but since GF's are decaying in time and since we are only interested in times $t \gg -\infty$, it is sufficient to consider $G(t, t')$ for $t, t' \in [t\_{min}, t\_{max}]$ where $t\_{min}$ is a numerical parameter, in which convergence has to be assured. 
We start the KBe time evolution at $t = 0$, thus providing initial conditions $G(t, t')$ for $ t, t' < 0$ from the numerical iteration of the equilibrium KBe's; see end of last section.

Time steps are performed using a $p$th order linear multistep method (predictor-corrector) in combination with Gregory integration \cite{fornbergImprovingAccuracy2021} for the history integrals, on a uniform time-grid $t_n = t\_{min} + n\delta t$; $\delta t$ denoting the step-size. The two-time arguments of the Greens functions $G(t, t')$ hence become matrices with dimension $N_t \times N_t$, where $N_t = (t\_{max} - t\_{min}) / \delta t$.

For first order electronic equations [\cref{eq:supp_final1,eq:supp_final2}] we use the $p$th order Adams-Moulten predictor-corrector method \cite{butcherNumericalMethods2016} and for second order bosonic equations [\cref{eq:supp_final3,eq:supp_final4}] we use the $p$th order Störmer's method \cite{hairerSolvingOrdinary2000}. The latter is a predictor-corrector scheme for differential equations only containing $\partial_t^2$, but no $\partial_t$. It is more accurate and numerically less expensive than mapping the second order equation onto two first order equations. The global integration error (full integration) for both $p$th order methods is $\order*{\delta t^{p+1}}$. The required computer memory for the KBe's scales as $\mathcal{O}(N_t^2)$, while the computational effort scales as $\mathcal{O}(N_t^3)$.

All data in this paper is generated with $p = 5$, i.e. an $\mathcal{O}(\delta t^6)$ scheme. Typical step sizes used in the integration are $\delta t = 0.02 - 0.05$, leading to matrix dimensions $N_t = 10.000 - 30.000$, depending on $t\_{max}$. Overall convergence of the results is most difficult to achieve at low temperatures and/or strong coupling $g$. For large couplings $g \geq 3$, a tiny $\delta t < 0.01$ is required to achieve numerical convergence, making the analysis in this regime numerically unfeasible. 

\paragraph*{\textbf{Scaling of the method} ---}
Performing an equilibrium time evolution with the KBe's presents itself as a sensitive consistency check for the KBe solver. Using the equilibrium (eq) GF's as the initial condition for $t<0$, we propagate the GF's to $t\_{max}$ using the KBe's. This is done without quench or time dependence in the Hamiltonian, i.e. with the equilibrium functionals $\Sigma[G]$. Since $G(t, t') = G(t - t')$ in equilibrium, the Gf's should not change along the propagation with the KBe's. We can thus compare the propagated GF's with the eq GF's at the end of the time evolution to measure the error of our numerical integration.

\begin{figure}[t]
    \centering
    \includegraphics{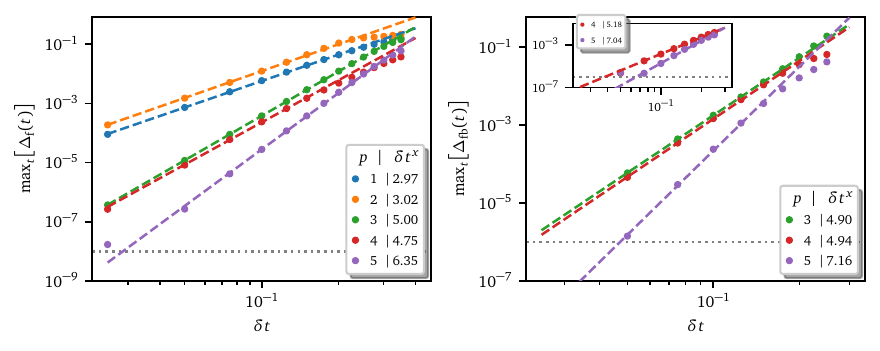}
    \caption{\textbf{Error scaling of equilibrium time evolution with KBe's}. Shown is the maximum error recorded during the time evolution. The expected scaling of the $p$th order methods is $\order*{\delta t^{p+1}}$. The dashed lines are linear fits to the $\log$-data and the slope of this fit is shown in the right column of the legend. The gray dotted lines indicate the estimated accuracy of the equilibrium solution.
    \textit{Left:} SYK model with $\beta = 10, \mu = 0.1, J = 1.2, K = 0.3$, see \cite{kuhlenkampPeriodicallyDriven2020} for definition of model and parameters. 
    The GF's are integrated from $t = 0$ to $t\_{max} = -t\_{min} = 100$.
    \textit{Right:} YSYK model with $\beta = 1.2, \mu = 0.0, g = 1.0$. The GF's are integrated from $t = 0$ to $t\_{max} = -t\_{min} = 250$. The inset shows the scaling for the superconducting YSYK model with $\beta = 27.03, \mu = 0.0, g = 0.6$ with the same numerical parameters as the normal state.}
    \label{fig:nkbe_kbe_scaling}
\end{figure}

Since the implementation that calculates the eq GF's and the KBe code are completely separate, this provides a great consistency check, not only the KBe code, but also for the code that calculates the equilibrium solution of the system. The scaling analysis of this equilibrium propagation is shown in \cref{fig:nkbe_kbe_scaling}. The results for the propagation of the SYK equilibrium solution are shown on the left (see \cite{kuhlenkampPeriodicallyDriven2020} for definition of the model) and for results for the normal state of the YSYK model on the right. The inset shows the scaling for the superconducting YSYK model. The YSYK model has not been considered in non-eq before, so that here the right panel presents a very sensitive consistency check for our novel implementation.

\section{Quench with Finite Ramping Time}



\begin{figure}[t]
    \centering
    \includegraphics{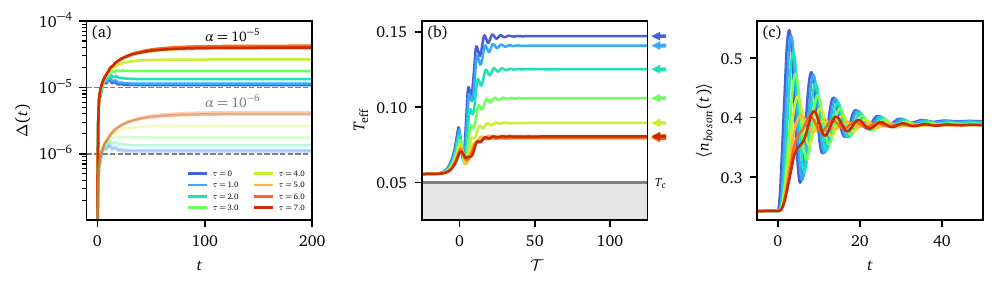}
    \caption{\textbf{Quenches across superconducting transition with finite ramping time} for $g = 0.75 \to 1.0$ with $\alpha = 10^{-5}$ (full lines) and $\alpha = 10^{-6}$ (desaturated lines; only distinguishable in the left panel). (a) The superconducting order parameter $\Delta(t)$ stays on the order of the symmetry breaking $\Delta(t) \sim \alpha$, implying that no superconductivity is induced. (b) After initial oscillations at early times, the effective temperature $T\_{eff}$ agrees with the instant temperature $T\_{inst}$ (color matched arrows), indicating thermalization at late times. Universally $T\_{eff}(t \to \infty) > T\_c$. (c) The bosonic ocuupation $\expval{n\_{boson}(t)}$ follows an analog behavior to the effective temperature, viz. oscillations at early times that relax in the long time limit.}
    \label{fig:supp_finite_ramp}
    \label[fig_a]{fig:supp_finite_ramp_delta}
    \label[fig_b]{fig:supp_finite_ramp_temp}
    \label[fig_c]{fig:supp_finite_ramp_nb}
\end{figure}

The inability to induce superconductivity following a quench across the phase boundary also persists when considering a finite ramp for the interaction quench. We study this for quenches from $g\_i = 0.75$ to $g\_f = 1.0$, where the finite ramping time $\tau$ is implemented as a linear ramp in $g_{ij,k} f(t)$ with
\begin{align}
    f(t) = g\_i \theta(-t) + \left(g\_i + \frac{g\_f - g\_i}{\tau} t \right) \theta(\tau - t) + g\_f \theta(t - \tau).
\end{align}
The exact dynamics for ramping times $\tau = 0, \dots, 7$, where $\tau = 0$ corresponds to  instant quenches, is shown in \cref{fig:supp_finite_ramp}. \Cref{fig:supp_finite_ramp_delta} plots the gap function $\Delta(t)$, \cref{fig:supp_finite_ramp_temp} the effective temperature and \cref{fig:supp_finite_ramp_nb} the bosonic occupation. A further increase in ramping time does not change the dynamics in any meaningful way, nor decrease the final effective temperature. 

Analog to the instant quenches, the system relaxes to the new equilibrium state with an oscillatory dynamics. While it is possible to decrease the final temperature after the quench by increasing the ramp time $\tau$, it does not seem to be possible to achieve $T\_{eff} < T\_c$, at least for the coupling considered here. The final temperature saturates for $\tau > 6$ [orange and red curves in \cref{fig:supp_finite_ramp_temp}].  We universally find $\Delta(t) \sim \alpha$ [saturated and desaturated; \cref{fig:supp_finite_ramp_delta}], implying that no transient superconductivity is induced.

\begin{figure}[h!]
    \centering
    \includegraphics{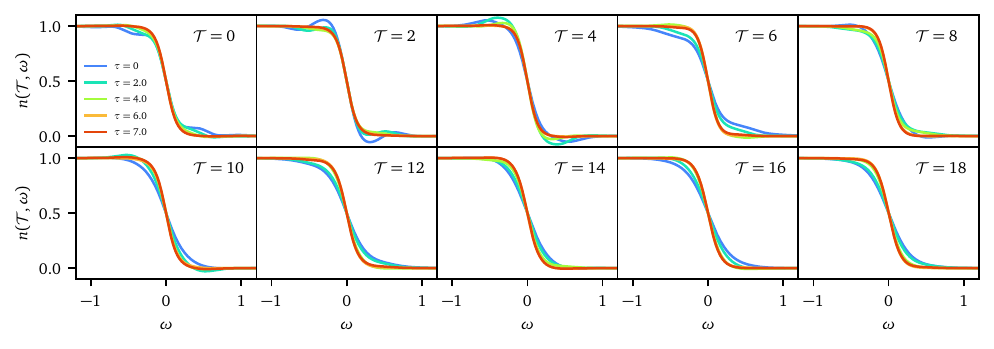}
    \caption{\textbf{Non-equilibrium ocuupation function $\boldsymbol{n(\mathcal{T}, \omega)}$ after quenches across superconducting transition} at fixed $\alpha = 10^{-5}$ for instant (blue) quenches and finite ramps. While instant quenches lead to strong non-equilibrium features in the non-equilibrium occupation function ($n > 1, n < 0$), these features gradually vanish as the ramping time increases.}
    \label{fig:supp_finite_ramp_wigner}
\end{figure}
At the largest ramping times $\tau \geq 6$, the system is `locally thermal' during and after the quench, as we illustrate in \cref{fig:supp_finite_ramp_wigner}, which shows the non-equilibrium occupation function $n(\mathcal{T}, \omega)$ for early and intermediate times. While for instant quenches and short ramps, the occupation function $n(\mathcal{T}, \omega)$ shows strong non-eq features ($n > 1$ or $n < 0$), these gradually decrease and vanish as the ramping times $\tau$ increases. At $\tau \geq 6$, the occupation function resembles as Fermi-distribution with time dependent temperature during the time evolution.

\section{Thermalization Time and Critical Slowing Down}

\begin{figure}[b]
    \centering 
    \includegraphics{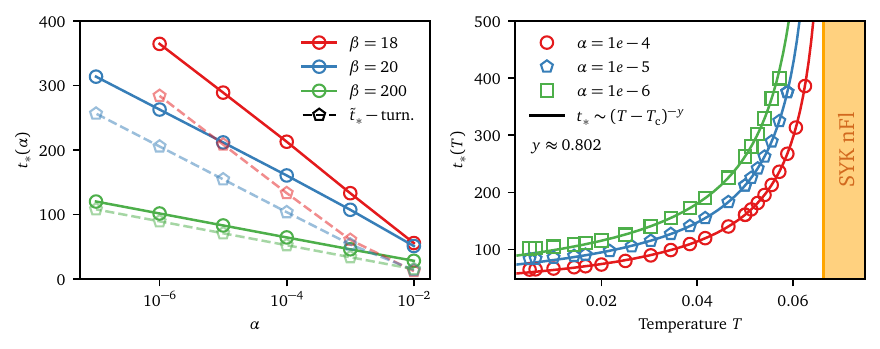}
    \caption{\textbf{Parametric dependence of thermalization time $\boldsymbol{t_*}$.} (a) $t_*(\alpha)$ as a function of the symmetry breaking strength $\alpha$ for various temperatures. Both thermalization $t_*$ (circle, full) and turning point $\tilde{t}_*$ (pentagon, dashes) follow a linear curve $t_*(\alpha) = t_{*,0} + C \log(1/\alpha)$, indicating that the relaxation is exponentially activated in $\alpha$. (b) Temperature dependence of $t_*(T)$. Close to $T\_c$ (normal state; orange) we observe a critical slowing down, i.e. $t_*(T) \sim (T - T\_c)^{-y}$ for which we extract the $\alpha$ independent exponent $y$ by fitting $T \geq 0.03$. For temperatures $T \ll T\_c$, the exact $t_*(T)$ deviates from this power-law form and saturates.}
    \label{fig:ysyk_parametric_dependence}
    \label[fig_a]{fig:ysyk_parametric_dependence_a}
    \label[fig_b]{fig:ysyk_parametric_dependence_b}
\end{figure}

In the main text we remarked on the strong $\alpha$ dependence of the thermalization time $t_*$. Its functional dependence given by 
\begin{equation}
    t_*(\alpha) = t_{*,0} + C \log(1 / \alpha),
\end{equation}
with temperature dependent constants $C, t_{*,0} > 0$, as we illustrate in \cref{fig:ysyk_parametric_dependence_a} (note the logarithmic x-axis). The turning-point time $\tilde{t}_*$, up to which the order parameter shows an exponential behavior, has exactly the same slope, but a different constant $ t_{*,0} \to \tilde{t}_{*,0}$. The dynamical onset of superconductivity is hence exponentially activated in the strength of the symmetry breaking seed. This is readily understood via the exponential dynamics at early times $\Delta(t) \sim \exp(\Gamma t)$. The system inherently produces an exponentially growing seed, a short time after the initial symmetry breaking occurs. Hence, only providing an exponentially bigger initial seed can modify this self-propelled dynamics in a meaningful way.

Another interesting question is the parametric dependence of $t_*$ on the initial temperature $T$ of the normal state, which we analyze in \cref{fig:ysyk_parametric_dependence_b}. Here, we observe a critical slowing down of the relaxation dynamics as we approach the critical temperature $T\_c$ (orange region), i.e. $\lim_{T \to T\_c}t_*(T) = \infty$, which is simply a different manifestation of $\lim_{T \to T\_c} \Gamma(t) = 0$ discussed in the main text \cite{hohenbergTheoryDynamic1977}. This feature is very intuitive, since the closer we get to $T\_c$, the smaller the `effective force' becomes, that pushes the system into a superconducting ground state. Right at $T\_c$, the system is critical so that the restoring force vanishes and the thermalization time diverges. 

\begin{SCfigure}[][t]
    \centering
    \includegraphics{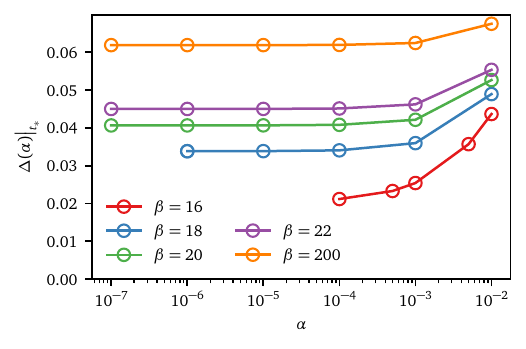}
    \caption{\textbf{Relaxed gap function $\Delta(\alpha)\vert_{t_*}$} at thermalization time $t_*$, for $g = 1.0$ for different temperatures. Close to the superconducting transition temperature ($T\_c \approx 1 / 15.1$) the relaxed $\Delta$ depends strongly on $\alpha$, while deep within the superconducting phase it is much less sensitive. Since we want the final state to be independent of $\alpha$ on the scale of the plots, we must decrease it as we approach $T\_c$. This, combined with the critical slowing down (see above), makes the numerical study of undercooled quenches close to $T\_c$ challenging.}
    \label{fig:supp_delta_alpha}
\end{SCfigure}

While the dynamics is exponentially activated in the symmetry breaking strength, the thermalized value $\Delta(t > t_*)$ as well as the relaxation rate $\Gamma(T)$ are $\alpha$-independent, for small enough $\alpha$. That is desirable, since $\alpha$ is supposed to represent an infinitesimally small perturbation to the system that kick-starts our relaxation process, but does not affect the final state after relaxation (at least on the scale of the plots). This in turn means, that as we approach $T\_c$, $\alpha$ needs to be become smaller for the final state to be $\alpha$-independent, since the thermal equilibrium value of $\Delta$ continuously decreases to zero for $T \to T\_c$. We illustrate this in \cref{fig:supp_delta_alpha}. This in combination with the critical slowing down makes the analysis for $T \sim T\_c$ numerically challenging.
%

\section{Undercooling at Different Coupling Strengths}

\begin{SCfigure}[][h!]
    \centering
    \includegraphics{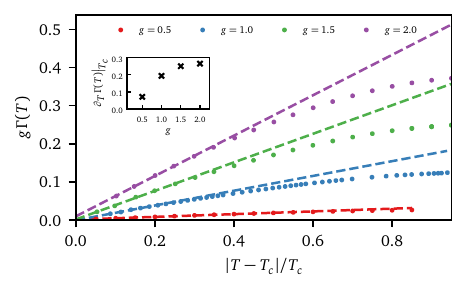}
    \caption{\textbf{Relaxation rate $\Gamma(T)$} extracted from early time dynamics $\Delta(t) \sim e^{\Gamma(T) t}$ at $\alpha = 10^{-5}$ and different couplings $g$. The main panel shows $g \cdot \Gamma(T)$ (for better visibility) together with linear fits (dashed lines) $\Gamma(T) \sim c_1|T - T\_c| / T\_c + c_2$ to the exact data (dots). The inset illustrates the $g$-dependence for the slope of the linear fit $c_1 = \partial_T \Gamma(T) \big|_{T \to T\_c}$. Note that $c_2 \approx 0$, indicating the critical slowing down. The dynamics is generally faster at stronger coupling $\partial_g c_1(g) > 0$, but like the critical temperature, saturates for $g \gtrsim 2$.}
    \label{fig:supp_rate_gs}
\end{SCfigure}

Here we comment on the undercooling dynamics at different coupling strengths $g$. The relaxation rate $\Gamma(T)$ extracted from the early time dynamics, is shown in \cref{fig:supp_rate_gs} for a variety of different couplings (note the rescaled y-axis for better visibility). Close to $T\_c$ the prediction from overdamped Ginzburg-Landau theory (dashed) agrees well with the exact results, while the curves universally deviate from this behavior for stronger undercooling, where they show a saturation behavior. We observe that the dynamics is generally faster at stronger coupling. Further, we find that the oscillatory order parameter dynamics is most pronounced at weak and intermediate coupling $g \lesssim 1$,  while the amplitude of the oscillations becomes strongly suppressed at strong coupling.
%

\section{Ginzburg-Landau Parameter Fits}

We analyze the temperature and coupling dependence of the Ginzburg Landau theory (GL) parameters, mentioned in the main text. We obtained these from fits of the Ginzburg-Landau $\Delta\_{GL}(t)$ to the exact order parameter dynamics. Our Ginzburg Landau theory is defined by the Free energy $F\_{GL} = a \Delta\_{GL}^2 + b \Delta\_{GL}^4$ and dynamical equation
\begin{align}
    \eta \partial_t^2 \Delta\_{GL}(t) + \delta \partial_t \Delta\_{GL}(t) = -\frac{\partial F\_{GL}}{\partial \Delta\_{GL}}, \qquad \Delta\_{GL}(t = 0) = \alpha
    \label{supp:gl_dynamics}
\end{align} 
which for $\eta \neq 0$ has to be solved numerically due to the non-linearity. Generally we find that the fitted parameters are very sensitive to small deviations in $\Delta(t)$, making an accurate numerical extraction difficult. The extracted values lead to $\Delta\_{GL}(t)$-curves, that on the scale of the plots, coincide with the exact $\Delta(t)$ dynamics, but one shouldn't put too much emphasis on the exact numerical values obtained here. 

This procedure can nonetheless be used to determine the general trends for the GL parameters. The fits are obtained using the \texttt{BFGS} routine, implemented in \texttt{Optim.jl} \cite{mogensenOptimMathematical2018}, on $\Delta(t)$ and $\partial_t \Delta(t)$ together with \texttt{DifferentialEquations.jl} \cite{rackauckasDifferentialequationsJl2017} for the numerical solution of the ODE's. 

\begin{figure} 
    \centering
    \includegraphics{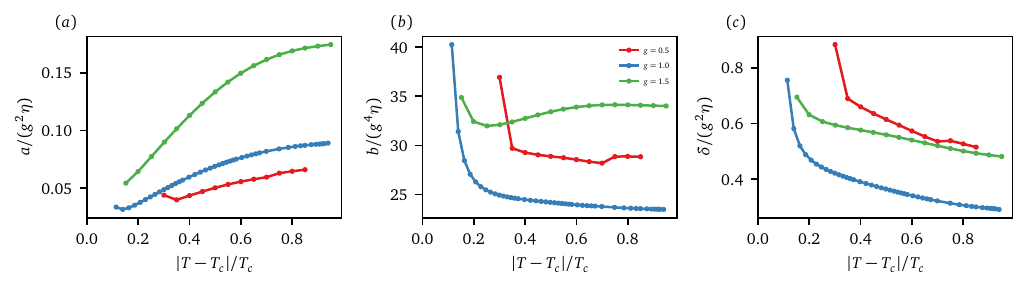}
    \caption{\textbf{Ginzburg-Landau parameters for undercooling protcol} at different couplings $g$. Axis are rescaled with the coupling strength $g$ and critical temperatures $T\_c$ for better visualization. While the exact numerical values are sensitive to numerical details, the general trends appear to be universal.}
    \label{fig:supp_glfits}
    \label[fig_a]{fig:supp_glfits_a}
    \label[fig_b]{fig:supp_glfits_b}
    \label[fig_c]{fig:supp_glfits_delta}
\end{figure}

The extracted parameters are shown in \cref{fig:supp_glfits} for different couplings $g$. Note that generally $a / \eta \to 0$ as $T \to T\_c$ as expected [\cref{fig:supp_glfits_a}]. In this limit $\delta / \eta \to \infty$ [\cref{fig:supp_glfits_delta}], implying that the dynamics becomes overdamped. We independently verified this by fitting GL theory with $\eta = 0$ to the exact numerical results close to $T\_c$. This leads to excellent agreement with the exact data.

We generally find larger damping $\delta$ in the stronger coupling regime, with the rough scaling $\delta \sim \mathcal{O}(g^2)$; curves in the figure become of the same order of magnitude when applying this rescaling. Hence, in order to observe the oscillatory order parameter dynamics at strong coupling, a stronger undercooling is necessary.
%

\bibliographystyle{apsrev4-2}
\bibliography{references_syk.bib}